\documentclass[preprint]{aastex}

\newcommand{\ms}{m~s$^{-1}$}
\newcommand{\eb}{\begin{equation}}
\newcommand{\ee}{\end{equation}}
\newcommand{\phmi}{\phantom{-}}

\shorttitle{Surface flows on the Sun}
\shortauthors{Makarov}
\begin{document}

\title{Variability of surface flows on the Sun and the implications for exoplanet detection} 
\author{Valeri V. Makarov}
\affil{NASA Exoplanet Science Institute, Caltech, 770 S. Wilson Ave.,
MS 100-22, Pasadena, CA 91125}
\email{vvm@caltech.edu}

\begin{abstract}
The published Mount Wilson Doppler-shift measurements of the solar velocity field 
taken in 1967--1982 are revisited
with a more accurate model, which includes two terms representing the meridional
flow and three terms corresponding to the convective limb shift. Integration of
the recomputed data over the visible hemisphere reveals significant variability
of the net radial velocity at characteristic time scales of $0.1$--$10$ years,
with a standard deviation of 1.4 \ms. This result is supported by independent
published observations. The implications for exoplanet detection include reduced
sensitivity of the Doppler method to Earth-like planets in
the habitable zone, and an elevated probability of false detections at periods of
a few to several years. 
\end{abstract}

\label{firstpage}
\section{Introduction}
The spectacular success of the exoplanet search program, resulting in the discovery of
over 300 planets and planetary system to date, has been achieved mostly through
the Doppler-shift technique. Indeed, spectroscopic detections constitute the
majority of known systems \citep{but, udr}, with the bias toward short-period,
massive planets probably due to the selection effects of this method. Undoubtedly,
there is much room for further progress with the Doppler technique \citep{egg},
with the accuracy of spectroscopic instruments steadily improving and now reaching
$\sim 1$ \ms\ and the sensitivity of telescopes extending toward fainter stars.
Combining precision photometric observations with radial velocity measurements
provides a range of important physical characteristics of transiting stars
invaluable for our understanding the physics and the origin of exoplanets.
The strategic goal of detecting rocky, habitable planets outside the Solar system
now seems to be coming within reach. As the instrumental precision steadily improved,
a growing attention has been paid to the intrinsic perturbations in the observable
parameters used in exoplanet detection. Stochastic, uncorrelated physical perturbations
increase the level of noise, making it difficult to achieve the threshold signal-to-noise
ratio, while possible cyclic processes in the host stars can mimic exoplanet signatures.
In particular, the rotating pattern of photospheric spots  \citep{saa} and irregularities in
the convective structure on the surface \citep{meu} can lead to an intrinsic scatter of radial
velocities of up to a few \ms\ for solar type G dwarfs. In the younger
Hyades, which are more magnetically active and rotate faster than the Sun, this
effect is magnified to $\sim 16$ \ms\ in standard deviation, as observed by
\citet{pau}. As the level of activity is not constant in solar-type stars, the
intrinsic radial velocity scatter is correlated with the magnetic cycle, opening
possibilities of more sophisticated spectroscopic analysis in order to mitigate these
difficulties. For example, the index of chromospheric activity is correlated with
the area of star spots (and hence, with the observed scatter) and can serve as
an indicator of magnetically induced cycles \citep{san00}. Careful selection of
target stars can further improve the prospect of detection of smaller planets with the Doppler technique. Recent simultaneous measurements of chromospheric activity and radial velocity
imply that K-type dwarfs may be significantly less variable than the Sun, and
the intrinsic RV jitter is less than 1 \ms\ \citep{san10}. Some stars older than 6 Gyr
and evolved off the main sequence have sharply reduced levels of magnetic activity
\citep{wri} compared to the Sun. In this paper, we consider another important source of
intrinsic radial velocity variation, related to the physical motion of the surface
layers of stars, which remained largely outside the scope of previous papers.

The surface layers of the Sun, where the spectroscopic lines are formed, are known
to be involved in a complex pattern of radial and tangential motion. Furthermore,
it is now an observational fact that this velocity field is not static. In this
paper, the series of Doppler measurements taken at the Mount Wilson Observatory is
revisited, and the published fitting model coefficients are transformed to a
different model of differential rotation, convective blueshift and meridional flow,
which may more adequately represent the reality (\S~\ref{data.sec}). The transformed
model is integrated to produce the net radial velocity of the Sun as a star (\S~\ref{rv.sec}).
The results are compared to other data on the variability of the main components of
the velocity field in \S~\ref{comp.sec}, and a good agreement is found. The impact
of the intrinsic variation of solar RV on the detectability of exoplanets, especially
within the habitable zone, are investigated in \S~\ref{impact.sec} by means of $\chi^2$
and spectral density periodograms, as well as by a planet detection experiment,
which results in two bogus planets. The relative importance of surface flows compared with
other sources of intrinsic RV perturbations are discussed in \S~\ref{dis.sec}.

\section{Mount Wilson data}
\label{data.sec}
An extensive set of Doppler velocity measurements taken at Mount Wilson during a period of
14 years, from 1967 through 1982, was published by \citet{how}. The full-disk magnetograms
in Fe I line 5250.2 \AA\ were fitted with the model\footnote{The instrumental drift term
has been omitted}
\begin{eqnarray}
V &=&(A+B\,\sin^2\phi+C\,\sin^4\phi)\sin\,\lambda\,\cos\,\phi\,\cos\,b_0+ \nonumber \\
&& (D+E\,\sin^2\phi+F\,\sin^4\phi)\sin^2\lambda\,\cos\,\phi\,\cos\,b_0 + G
\label{abc.eq}
\end{eqnarray}
with $\phi$ the solar latitude, $\lambda$ the longitude (heliocentric angle from the central
meridian), $b_0$ the solar latitude of disk center, $G$ an arbitrary zero-point
velocity, and $A$, $B$, $C$, $D$, $E$ and $F$ free fitting parameters. The first
group of terms with coefficients $A$, $B$, and $C$ represents differential
rotation, with $A$ denoting the equatorial velocity. Since the terms of differential
rotation are odd functions of $\lambda$, their net contribution to the integral
radial velocity is zero, and we are not concerned with them in this paper.
The second group of terms with $D$, $E$ and $F$ can be interpreted as the
antisymmetric component of the differential rotation. These terms are
even in $\lambda$ and result in a net
contribution to the radial velocity. However, because the observations are differential
and the term $G$ is arbitrary, the total offset of velocity is indeterminate,
and only the variation of these terms in time can be considered.

The integral radial velocity as measured by a distant observer is
\eb
\bar V=\frac{1}{2\pi}\,\int_{-\frac{\pi}{2}}^{\frac{\pi}{2}}d\lambda
\,\int_{-\frac{\pi}{2}}^{\frac{\pi}{2}}(\alpha_0 +\alpha_1\cos\,\phi\,\cos\,\lambda
+\alpha_2\cos^2\,\phi\,\cos^2\,\lambda) \, V\,\cos\,\phi\,\cos\,\lambda\,\cos\,\phi\,
d\,\phi.
\label{int.eq}
\ee
In this formula, $\alpha_0$, $\alpha_1$
and $\alpha_2$ are the coefficients of solar limb darkening. Adopting $\alpha_0=0.30$,
$\alpha_1=0.93$ and $\alpha_3=-0.23$, the resulting net radial velocity is
\eb
\bar V =0.1005\,D+0.01841\,E+0.007417\,F.
\label{barv.eq}
\ee
Using the tabulated data in \citep{how}, we compute a radial velocity curve, which shows
significant variations of 4.95 \ms\ in standard deviation (STD). However, this result
is likely overestimated, for the following reasons.

As was pointed out by \citep{snoul,snoho}, the terms of differential rotation $A$, 
$B\,\sin^2\phi$ and $C\,\sin^4\phi$ are not orthogonal on a half-sphere,
resulting in considerable correlations between the coefficients determined
from regularly sampled data. They suggested to use Gegenbauer polynomials
in $\sin\phi$ to orthogonalize the terms. The first thus orthogonalized coefficient
corresponding to the principal component of solar rotation, is $\hat A = A + B/5 +
3\,C/35$. Note that the coefficients in this relation are closely in
the same proportion (1 : 0.2 : 0.085) as the coefficients in Eq.~\ref{barv.eq},
(0.1005 : 0.01841 : 0.007417). This coincidence may be interpreted as
a real net variation of the integral velocity (which can not be captured
by the odd $A$, $B$ and $C$ terms) translating into the non-orthogonal
$D$, $E$ and $F$ terms according to their covariances, since all the differential measurements
are brought to a single zero-point. This hypothesis is supported by the actual
negative correlation of the three terms in Eq.~\ref{barv.eq}, whose variances add up
to a larger number than $4.95^2$.

\begin{figure}[htbp]
\plotone{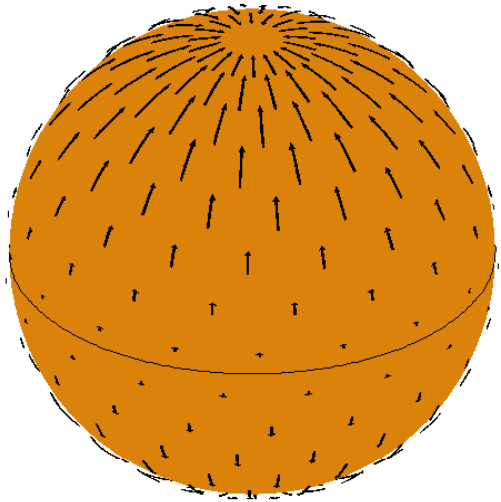}
\caption{The velocity field of solar meridional flow.
The net motion is directed from the equator toward the poles,
vanishing at the equator and the poles.} 
\label{sunV.fig}
\end{figure}
Thus, a real variation of the surface velocity field was probably
exaggerated in the model (\ref{abc.eq}) because the $DEF$ terms  are internally correlated. 
Besides, these terms poorly describe the real velocity field. Indeed, the functional 
form of $DEF$ terms 
corresponds to a
pattern of zonal stretches on the surface, which remain fixed with respect to
the central meridian despite the general rotation of the Sun. Clearly, any
longitudinal perturbations of rotation velocity can not remain static with
respect to the observer on time scales longer than one rotation period. The correct
physical explanation to the symmetric pattern of Doppler velocities observed
at Mount Wilson was given by \citet{lab}. The mysteriously looking zonal pattern
of radial velocity is caused by two separate physical phenomena
on the Sun: the axisymmetric convective blueshift (often called the limbshift)
and the meridional surface flow symmetric around the equator to first approximation.

The model of meridional flow considered by \citet{lab} assumed a constant surface velocity
directed toward the poles. We replace it with a more adequate model suggested by 
\citet{hat}:
\eb
V_{\rm M}=\mu_2 P_2^1(\sin\,\phi)+\mu_4 P_4^1(\sin\,\phi),
\ee
where $P_l^1$ are associated Legendre polynomials of degree $l$ and order 1. The free fitting
parameters $\mu_2$ and $\mu_4$ are to be determined from observations. The Legendre
polynomials are orthogonal on the range of $\sin\,\phi$, which drastically simplifies
the subsequent analysis. We also replace a polynomial expansion for limbshift with
the more sophisticated model from \citep{hat1}:
\eb
V_{\rm B}=\sum_{k=1}^3\beta_i P_k^*(1-\cos\,\rho),
\ee
where $P_k^*$ are shifted Legendre polynomials orthogonal on $[0,1]$, $\rho$ is the
central angle ($\cos\,\rho=\cos\,\phi\,\cos\,\lambda$), and $\beta_i$ are free
fitting coefficients. 

Thus, the new model, which more faithfully represents the observed velocity field is
\begin{eqnarray}
V &=&(A+B\,\sin^2\phi+C\,\sin^4\phi)\sin\,\lambda\,\cos\,\phi\,\cos\,b_0+ \nonumber \\
&& (\mu_2 P_2^1(\sin\,\phi)+\mu_4 P_4^1(\sin\,\phi))(\cos\,\lambda\,\sin\,\phi\,\cos\,b_0
-\cos\,\phi\,sin\,b_0)+V_{\rm B} + G.
\label{new.eq}
\end{eqnarray}
We carried over the $ABC$ terms of differential rotation (rather than using Gegenbauer
polynomials) to simplify subsequent transformations.

Our task is now to recompute the observed parameters of the velocity field using
the published $ABCDEF$ parameters. The fitting procedure is a linear least-squares problem
and the transformation problem can be expressed in matrix notation
\eb
\left[ {\bf \Lambda}\;{\bf \Phi}\right]\,{\bf \tilde x}=
\left[ {\bf \Lambda}\;{\bf T}\right]\,{\bf \bar x} + {\bf \epsilon},
\ee
where ${\bf \Lambda}$ is the matrix of differential rotation (the $ABC$ terms),
${\bf \Phi}$ is the matrix of the symmetric stretch (the $DEF$ terms), {\bf T} is
the matrix of the new meridional flow and limb shift terms, ${\bf \tilde x}$
is the vector of free parameters of the old model, ${\bf \bar x}$ is
the vector of free parameters of the new model, and ${\bf \epsilon}$ is
random noise. It can be shown that in a least-squares solution,
\eb
{\bf \bar x}={\bf T}^\dagger \left[ {\bf \Lambda}\;{\bf \Phi}\right]\,{\bf \tilde x},
\ee
where ${\bf T}^\dagger=({\bf T}^{\rm T}{\bf T})^{-1}{\bf T}^{\rm T}$ is the pseudoinverse
of ${\bf T}$. The matrix and inner vector products here are calculated through
integration of the element products over the visible disk with the $\cos\,b$
weight. The inner products of the new model terms and the $ABC$ terms are all zero
(in other terms, these functions are orthogonal). Therefore, the $ABC$ coefficients
remain unchanged in this transformation, and the new parameters depend only on
the $DEF$ coefficients.

After some toil with integration and matrix inversion, one obtains
\eb
\left[\matrix{\mu_2\cr \mu_4\cr \beta_1\cr\beta_2\cr\beta_3}\right]=
\left[\matrix{-0.766 & -0.252 & -0.119 \cr \phmi 0.482 & \phmi 0.094 &\phmi 0.018 \cr
\phmi 0.345 & \phmi 0.108 &\phmi 0.053 \cr \phmi 0.012 & \phmi 0.030 &\phmi 0.026 \cr
-0.106 & -0.030 & -0.010}\right]\cdot \left[\matrix{D\cr E\cr F}\right]
\ee

\section{The solar RV variation}
\label{rv.sec}
\begin{figure}[htbp]
\plotone{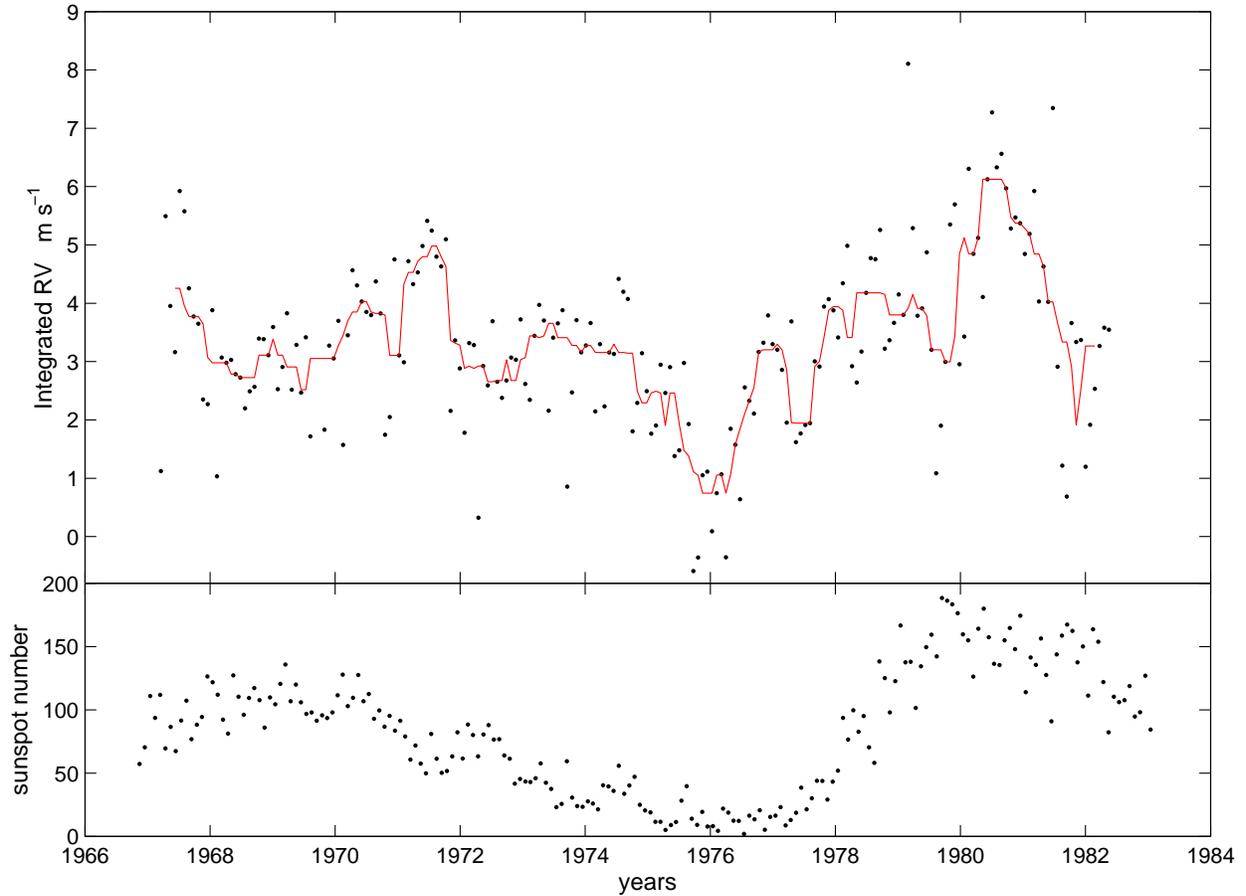}
\caption{The integrated radial velocity of the Sun seen equator-on
reconstructed from the Mount Wilson Doppler measurements (upper panel). A running median of
7 individual data points (one per Carrington rotation) is drawn with the
red line. The zero-point is arbitrary. In the lower panel, the sunspot number during
the same period is shown for reference.} 
\label{sunRV.fig}
\end{figure}

Analogous to Eq.~\ref{int.eq}, the new model of the solar velocity field can be
integrated over the visible hemisphere, to yield:
\eb
\bar V =-0.157\,\mu_2+0.025\,\mu_4-0.295\,\beta_1+0.044\,\beta_2+0.004\,\beta_3.
\label{barv.eq}
\ee
This result is valid only for the Sun seen edge-on (inclination $90\degr$), although
the limb shift is probably independent of inclination. {\it The five terms of the
new model are derived from only three terms of the old model, hence they are
strongly correlated because of the projection in the function space.}
The reconstructed radial velocity curve is shown in Fig.~\ref{sunRV.fig}.
The original data in \citep{how} are sampled by Carrington rotations, therefore,
each data point represents the mean over 27.2753 days. Monthly variations
of a few \ms\ are common, but so are longer-term variations on time scales of
years. A running median shown with the red line indicates that the long-term
radial velocity of the Sun grew by $\simeq 7$ \ms\ between 1976 and 1981.The
reconstructed RV curve has a standard deviation of 1.44 \ms~. The long-term
behavior of solar RV is strongly correlated with the magnetic cycle (cf.
the monthly sunspot numbers in the lower panel of Fig.~\ref{sunRV.fig}).
Radial velocity variations on time scales of several months to 1 year have
a strong impact on the detectability of habitable planets around solar-type
stars, as discussed in \S~\ref{impact.sec}. We have to carefully investigate if
these results are consistent with other independent data on the Sun.

\section{Comparison with other data}
\label{comp.sec}
Noting that the low-order terms of meridional flow ($\mu_2$) and limb shift
($\beta_1$) propagate with the largest weights into the integral velocity
in Eq.~\ref{barv.eq}, we have to consider what is known about the
variability of these physical effects. The most direct comparison comes
from the Doppler observations with the Global Oscillation Network
Group (GONG). \citet{hat} present the results for the main component of the
meridional flow over some 130 days in 1995 (their Fig. 1, top panel),
corresponding to $P_2^1$ (Legendre polynomial of degree 2 and order 1).
These coefficients are denoted $\mu_2$ in Eq.~\ref{barv.eq}. The flow
velocity varies between 0 and 50 \ms, which for an edge-on Sun translates
to a peak-to-peak amplitude of 8 \ms. Quasi-periodic cycles of 10--20
days are evident in the GONG data, as well as longer-term variations
on a few months, which are expected to change the integral velocity by
roughly 3--4 \ms. The amplitudes of the higher-degree terms $P_4^1$ and $P_6^1$
are found to be much smaller ($-1.6\pm0.5$ and $-1.3\pm0.1$ \ms), so their
contribution to the net radial velocity is negligible in the context of this
paper.

On longer time scales, the main $l=2$ component of meridional circulation
was investigated by \citet{hat1}, also based on the GONG Doppler measurements.
The flow velocity grew from $\sim 20$ \ms\ in the summer of 1992 to almost 100 \ms\
by the end of 1993, then dropping to small negative values in the summer of 1994,
indicating a brief episode of reversal. After that, it gradually returned to
$\sim 20$ \ms\ by the mid 1995. From Eq.~\ref{barv.eq}, the overall radial
velocity should have changed by 16 \ms\ in the first half of 1994. Unfortunately,
our data in Fig.~\ref{sunRV.fig} cover an earlier time interval, and can not be
directly compared with the GONG results. But the scale of variability in our
reconstruction is absolutely consistent with the newer, and more accurate data.

In the same paper, \citet{hat1} presents a reconstruction of the history
of the three terms of convective blueshift, denoted $\beta_1$, $\beta_2$
and $\beta_3$ in this paper. Only the first term appears to be a
significant contributor to the net velocity, with a quasi-sinusoidal
variation of $\sim 30$ \ms\ on a period of 1.5 yr and a short-term scatter of similar magnitude.
The corresponding range of radial velocity is $0.295\cdot 30/\sqrt{3}=5$ \ms.
Little correlation was noted between the meridional flow and the limb shift
on the GONG data.

The meridional flow traced by the motion of magnetic features in the photosphere
from a 26-year set of Mt. Wilson magnetograms appears to be somewhat smaller
in amplitude and more complex in the equatorial zone, with strong equator-ward
features \citep{snod}. The maximum rate is $\sim13$ \ms, and the pattern is
strongly correlated with the solar cycle. A detailed picture of the
solar cycle-related variation in the structure of the
meridional flow is shown in \citep[][their Fig. 2]{ulb}. The poleward components of the meridional
circulation are the strongest when the solar activity is at its maximum, in
agreement with the reconstructed data in Fig.~\ref{sunRV.fig}. \citet{giz}
used MDI full-disk Doppler images for the period 1996-2002 and measured the advection
of the supergranulation pattern, which is probably consistent with the tracing
of magnetic features. They detect only poleward motions reaching 13 \ms\ in the
antisymmetric part of the field averaged in 1 yr intervals. The peak-to-peak
variation of the antisymmetric component is 7 \ms. This is again in agreement
with the magnitude of RV variation derived in this paper.

With the abundance of high-quality data, the analysis procedures are very
complex, which accounts for some lingering disagreement between different
published results on the solar velocity field. One of the most rigorous and comprehensive
models was presented by \citet{hat87}, including differential rotation, torsional
streams, meridional
circulation, the limb shift, supergranules and giant cells. The surface velocity
field is represented with vector spherical harmonics, which are orthogonal on a
unit sphere. Although this much desired orthogonality is compromised by
the availability of only one hemisphere, inclination and projection effects
(as far as Doppler data are concerned), the various scales of poloidal and toroidal
motions are naturally separated in the model through vector spherical harmonics
of different degrees and orders. For example, the three main components of
differential rotation are represented by the toroidal harmonics of 
zero order ${\bf T}_1^0$ (solid body
rotation), ${\bf T}_3^0$ and ${\bf T}_5^0$. The limb shift terms are correlated (non-orthogonal)
to the meridional flow terms to such extent, that their separation requires
a full-cycle iteration \citep{hat91}, especially if the aim is to reconstruct
the {\it absolute} velocity field. This seems to be the only way to infer
the main term of limb shift (the value at the center of the disk), which is
$-540$ \ms\ according to Hathaway. Thus, the variable component of the limb shift
is only $\approx 1\%$ of the constant component.

Similarly, the large-scale meridional flows are dwarfed by the supergranulation
pattern, which has typical velocities of 300 to 400 \ms\ \citep{hat00}. The
high-degree supergranulation is of little interest in the context of this paper,
because it only contributes to the high-frequency variation of the integral
radial velocity, whereas we are interested in what happens with solar radial
velocity on the time scales of months to years. But the lower-degree supercells
are of interest, because they rotate with the Sun, indicating physical flows
in the photosphere. A rotating spherical harmonic term of order $m$ will produce
a periodic signal in the integral velocity with a period of $P_{\sun,{\rm rot}}/m$.
The power spectrum of velocity field is peaked at degrees $l\sim 120$, which
correspond to the spatial scale of supergranules. Therefore, rotating
supergranules will cause a net variation in radial velocity of a few \ms\ on
time scales of a few hours.

\section{The impact on exoplanet detection}
\label{impact.sec}
Exoplanets are found in Doppler measurement data through the periodic perturbations
caused by the orbital motion of the host star around the system barycenter.
To a good approximation, the radial velocity of the star is the sum of a constant
term (the velocity of the barycenter) and a number of Kepler motions. For low-eccentricity
orbits, such as those in the Solar System, the Kepler velocity signal is almost
perfectly sinusoidal. Detecting one or several sinusoidal variations in
high-cadence Doppler measurements superimposed with a constant offset is
fairly straightforward, if the observational noise is random, uncorrelated,
and sufficiently small compared to the expected signal. 

\begin{figure}[htbp]
\plotone{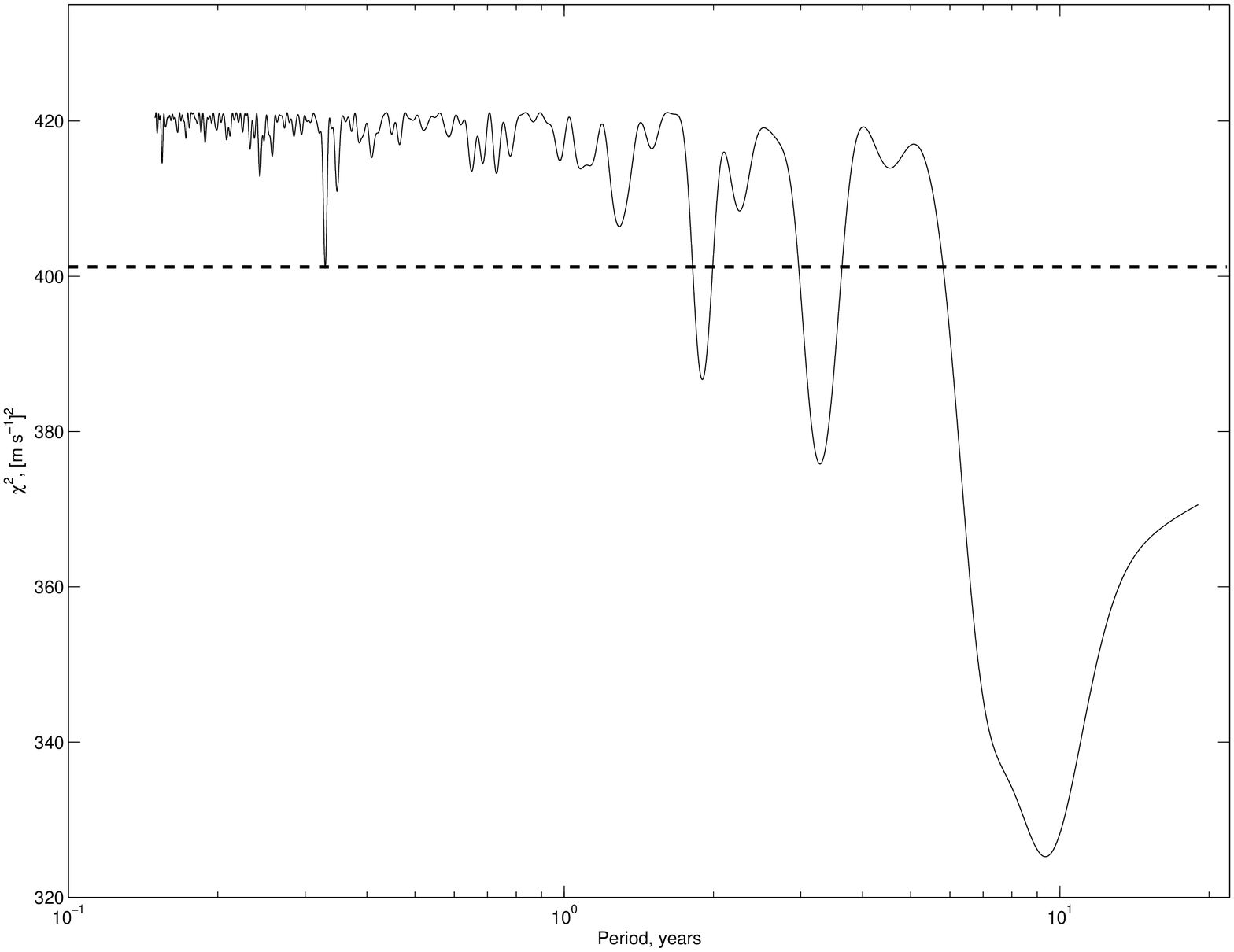}
\caption{$\chi^2$-periodogram of the reconstructed solar radial velocity curve
in Fig.~\ref{sunRV.fig}. The dashed horizontal line indicates the threshold
$\chi^2$ corresponding to a false alarm probability of 1\% for the first
planet detection.} 
\label{sunchi2.fig}
\end{figure}

An additional source of noise or perturbation from a real physical process
on the star makes this detection significantly more difficult,
or in some cases, ambiguous. To quantify the impact of the `surface flow noise' on the detectability of
planets, a generic planet detection algorithm was applied to the data
derived in \S~\ref{rv.sec}. No additional noise was added to the individual
RV points. The procedure starts with a periodogram analysis and a subsequent
estimation of the confidence of the most prominent sinusoidal variations.
Fig.~\ref{sunchi2.fig} shows the resulting periodogram of the reconstructed
solar radial velocities. Instead of the traditional spectral power, the
residual $\chi^2$ is plotted for a high-cadence grid of periods. The
$\chi^2$ statistics is preferred, because the confidence level of
the detected minima is rigorously computed via the cumulative probability
function of the $F$-statistics with $\{n_{\rm obs}-3,2\}$ degrees of freedom.
This confidence $F$-test is analogous to the estimation of FAP (false alarm
probability) by \citet{cum}, with the number of independent frequencies
corresponding to the frequency search interval. Each dip in this curve
which corresponds to a value of $F$ surpassing a given confidence threshold,
can be considered a planet detection. Setting the confidence threshold at
99\%, this periodogram analysis results in the detection of two bogus planets
with orbital periods $P_1=9.35$ yr (mass $26 M_\earth$) and $P_2=6.35$ yr
(mass $15 M_\earth$). The former false detection is apparently triggered
by the long-term variation of the meridional flow with the solar cycle, and
the latter may also be related to some poorly understood cyclicity of magnetic
activity, because a sinusoidal variation of similar frequency emerges in
the Total Solar Irradiance (TSI) data. 

\begin{figure}[htbp]
\plotone{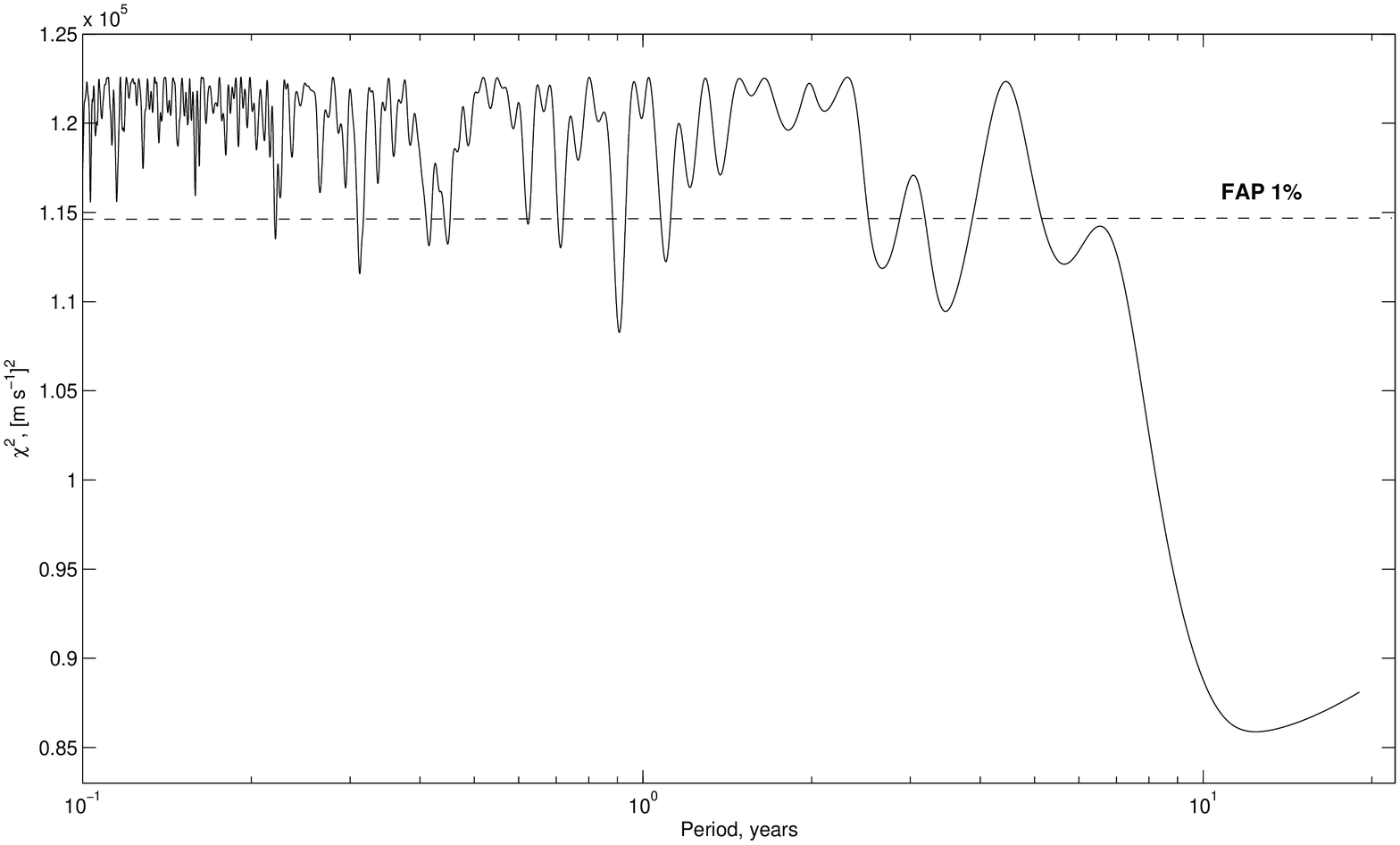}
\caption{$\chi^2$-periodogram of the RV measurements of 55 Cnc,
which were used to detect the 5 exoplanets reported in
the literature.  The dashed horizontal line indicates the threshold
$\chi^2$ corresponding to a false alarm probability of 1\% for the first
planet detection.} 
\label{55chi2.fig}
\end{figure}

This numerical experiment indicates that the magnetic cycle-related variations
in Sun-like stars can in some cases be confused with long-period planets. It is therefore
of interest to compare the reconstructed solar periodogram with accurate RV data
for a planet host. Fig.~\ref{55chi2.fig} shows a $\chi^2$ periodogram for the
star 55 Cnc with high-resolution spectroscopic measurements from \citep{mar, fis}.
The total span of these measurements is $18.3$ yr, and the formal single
measurement error progressed from $\sim9$ \ms\ during the first 6 years (Lick
observations) to between 3 and 5 \ms\ over the later period (Keck observations). 
The star is a K0/G8 dwarf, slowly rotating and chromospherically inactive
($\log R'_{\rm HK}=-4.84$). It is a close, albeit somewhat smaller, solar analog.
Five planets have been reported in the literature \citep{fis}, with orbital
periods from $2.8$ d to $14$ yr. Fig.~\ref{55chi2.fig} depicts a portion of
the  $\chi^2$-periodogram
computed from the weighted RV data for the same range of periods as Fig.~\ref{sunchi2.fig}.
Our confidence estimation is consistent with the FAP thresholds computed in the
cited papers, where a different technique was used (Monte-Carlo trials on randomly
scrambled data).
Except for a number of pronounced high-frequency dips, some of which correspond
to the detected planets, the structure of the periodogram is quite similar to
that for the Sun. In particular, the lowest dip in this part of the periodogram
centered on 14 yr at the long-period end, corresponding to the planet 55 Cnc d,
is reminiscent of the $9.35$-yr feature for the Sun, which is quite likely caused
by the solar cycle. This similarity calls for caution in the interpretation of
long-period signals from stars like 55 Cnc.

Extrapolating the observational results presented in \S~\ref{data.sec} to other
stars requires caution. The Mount Wilson data are based on a single Fe line, whereas
an RV curve for exoplanet detection is derived from a large number of weak lines
across the visible spectrum. There are good reasons to assume, however, that the
dominating $\mu$-terms of the meridional flow in Eq.~\ref{new.eq} are not wavelength-dependent.
The existing models of motions in the Sun (still somewhat uncertain) imply large-scale
and deep-rooted dynamical structures of global character. For example, \citet{how87}
presented a model of only several gigantic longitudinal supercells (or rolls), which
may evolve or drift with the magnetic cycle. In this interpretation, the large-scale
meridional flows are quite deep. The case with convective limbshift,
which is wavelength-dependent, is perhaps more complicated. The blueshift is not
the same in different lines, but it seems plausible that the long-term changes 
due to the 11-yr and other known magnetic cycles are caused by a slower or faster convective
motion on the integrated disk, thus resulting in a uniform shift of line bisectors across
the spectrum. Systematic changes in line shapes can be expected due to magnetic cycles as well.
Recently, \citet{san10} investigated a sample of eight late G or early K dwarfs of
low magnetic activity and compared precision RV measurements with simultaneous 
estimates of chromospheric activity. They could not find any clear correlation and
concluded that any RV variations induced by magnetic cycles on these stars are
unlikely to be much greater than 1 \ms\ . Therefore, it is possible that many K dwarfs
are less subject to the adverse effects of variable surface flows than more solar-like
stars.

Most of the currently known or suspected exoplanets have masses of the Solar System giants
(or larger), and short-period orbits \citep{but,udr}. A considerable effort is under way
to improve the sensitivity of the Doppler instruments and data calibration to the
levels sufficient to detect Earth-like, rocky planets within the habitable zones
of main sequence stars, where water can be present in the liquid state. The required
single measurement error due to photon noise and instrument imperfection should
be $\lesssim 0.1$ \ms\ to achieve this goal. One should consider the possibility
that the surface flow and convective blueshift perturbations are in fact much larger
than 0.1 \ms\ in the range of habitable orbits too. Using the reconstructed solar
RV curve in Fig.~\ref{sunRV.fig}, I computed the spectral power density (i.e., the
amplitude of the best-fitting sinusoid as a function of frequency) for a range of
periods $0.5$ through $1.5$ yr, roughly corresponding to the solar habitable zone.
The power density turned out to be quite flat at $\simeq 1$ \ms. This result is of modest value
because it only sets the upper
bound for the actual power spectrum of RV variations of the Sun, since the
contribution of observational error in the Mount Wilson data is not known. Ultra-precise, long-term RV observations of stable solar-type
stars will be required to confirm that the intrinsic RV variations come up to
such values. We note that a $3\sigma$ detection of Earth in a periodogram 
requires the perturbation power density to be less than 0.03 \ms.

\section{Discussion}
\label{dis.sec}
Variable surface flows are only one of the known physical processes on the Sun
that can change the integrated radial velocity. They appear to be long-term
in nature, and therefore, may be the main obstacle to detecting Earth-like
habitable planets with ultra-precise Doppler measurements. The main pulsation
modes of solar-type stars ($p$-modes) have periods of 10--20 min, and can be
successfully suppressed in observations by taking longer exposures. The supergranulation
pattern, discussed in \S~\ref{comp.sec}, should produce variations of up to a few
\ms\, but their time scale is several hours. Rotation of stars and the non-uniform
distribution of surface brightness due to photospheric spots and plages
is probably the main source of RV perturbations on the time scales
up to 50 d. The impact of spots has been investigated in numerous papers,
and the latest estimates suggest relatively small, but non-negligible,
dispersion for the Sun of $\sigma_{\rm RV} \simeq 0.4$ \ms\ \citep{mak}. As was
shown by \citet{pau} for the Hyades, the photometric and the RV jitter
of more active stars than the Sun are strongly correlated. Extrapolating the
empirical relation found in that paper, one would expect a jitter of 
$\sigma_{\rm RV} = 1.7$ \ms\ for the Sun. The large discrepancy between these
estimates is probably related to the different morphology of photospheric inhomogeneities
in the Hyades stars. Dark spots are the dominating magnetic features on active
and young stars, whereas the contribution of bright plages on older, solar-type
stars tends to match the impact of spots, or even to prevail \citep{loc}. 
Besides, active stars often have only one or two giant long-lived spots on the surface,
resulting in a strong rotational modulation of the light curve, whereas the surface
of an active sun is usually marked with a few spot groups at a time, fairly uniformly
distributed in longitude. \citet{meu} estimated the combined impact of sunspots and
plages at $\simeq 1$ \ms\ in RV amplitude, and concluded that it should be confined
to periods less than 100 d.
Our study suggests that on the time scales
$0.6$--$1.4$ yr, characteristic of orbits within the habitable zones, slowly evolving
surface flows and the distribution of the convective blueshift is the major
source of confusion and error.

Our conclusions are in agreement with the observations of solar radial velocity
by \citet{mkm}. These authors observed the solar light reflected from the Moon
for 5 years nightly in violet absorption lines and determined an upper limit
of 4 \ms\ for the overall dispersion of radial velocities. The periodogram of
RV variations (their Fig. 4) shows four peaks rising above the $3\sigma$ threshold,
all in the long-period part of the spectrum. Three of them may be of instrumental
origin, but the unresolved peak at period longer than 3 yr appears to be genuine
and is called "intriguing" in the paper. The two bogus planets found in the
reconstructed RV data in this paper (\S~\ref{impact.sec}) have periods $6.35$
and $9.35$ yr. Therefore, it is possible that \citet{mkm} already detected
long-term variations in the RV of the Sun as a star. \citet{meu} pointed out
that the effects of convective blueshift, which is included in our RV model (\S~\ref{data.sec}),
may be smaller in the deep violet lines than in the redder lines that are normally
used for planet search.

Our final note is that the model of meridional flow depicted in Fig.~\ref{sunV.fig}
is a simplified representation of the actual fine structure of the velocity field.
By necessity, the model for the Mount Wilson data is coarse, where the contribution
of convective blueshift and meridional flow can not be decoupled. The former is
especially problematic, because it is wavelength-dependent, and uncalibrated
changes in the spectrograph setup can bring about long-term systematic errors \citep[cf.][]{ulr}.
The low-order polynomials used to fit the surface velocity field (\S~\ref{data.sec})
only reveal the underlying, very much smoothed structure of the actual distribution
of tangential and radial motion, characterized by smaller spatial scales and 
larger amplitudes. For example, the local flows are known to converge toward
active regions, reaching 50--100 \ms\ \citep{giz4}, and these local inflows may
be responsible for the temporal variation of the overall velocity field on the
time scales of supergranulation (8 hours) and the solar cycle (11 years). Thus,
the temporal behavior of the velocity field may be in part the average of many
stochastic components, and in part the evolution of the global magnetic field \citep{ulb},
both still poorly understood for the Sun. Therefore, it is doubtful that a good
diagnostics can be devised for other stars to separate the physical motion of the surface from
the signatures of small planets.

\acknowledgments
The research described in this paper was carried out at the Jet Propulsion 
Laboratory, California Institute of Technology, under a contract with the National 
Aeronautics and Space Administration.

\end{document}